\documentclass[10pt,a4paper]{revtex4}
\usepackage{graphicx}

\begin{document}
\title{Uncertainties of Antiproton and Positron Spectra from B/C Data and mSUGRA Contributions for Clumpy Halos}

\author{A.M. Lionetto}
\affiliation{INFN, Sezione di Roma II, via della Ricerca Scientifica,Roma, Italy}
\affiliation{Dipartimento di Fisica, Universit\`a di Roma "Tor Vergata", via della Ricerca Scientifica, Roma, Italy}
\author{A. Morselli} 
\affiliation{INFN, Sezione di Roma II, via della Ricerca Scientifica,Roma, Italy}
\affiliation{Dipartimento di Fisica, Universit\`a di Roma "Tor Vergata", via della Ricerca Scientifica, Roma, Italy}
\author{V. Zdravkovi\'c}
\affiliation{Dipartimento di Fisica, Universit\`a di Roma "Tor Vergata", via della Ricerca Scientifica, Roma, Italy}

\begin{abstract}
We have studied the variation of $e^+$ and $\bar p$ top of the atmosphere spectra due to the parameters uncertainties of the Milky Way geometry, propagation models and cross sections. We have used the B/C data and Galprop code for propagation analysis. We have also derived the uncertainty bands for subFe/Fe ratio, H and He. Finally we have considered a neutralino induced component in the mSUGRA framework. We conclude that even in the case of diffusion convection model for the standard propagation, where secondary spectra gives a good fit of the data, SUSY contribution can not be excluded up to now due to the overall uncertainties.
\end{abstract}

\maketitle

\section{Production and Propagation of Cosmic Rays in the Milky Way}

We have chosen Galprop \cite{SM1} as a public code for the treatment of propagation of all cosmic rays (CR) together. Our scope has been to determine the total uncertainties in the calculation of $e^+$ and $\bar p$ top of the atmosphere spectra due to the uncertainties of geometrical and propagation parameters and cross sections.  Here we give very short description of processes included in propagation equation

\begin{eqnarray}
\frac{\partial \psi ({\mathbf r},p,t) }{\partial t} 
&=& q({\mathbf r}, p)+ \nabla \cdot ( D_{xx}\nabla\psi - {\mathbf V_c} \psi )
+ \frac{d}{dp}\, p^2 D_{pp} \frac{d}{dp}\, \frac{1}{p^2}\, \psi \nonumber\\
&-& \frac{\partial}{\partial p} \left[\dot{p} \psi
- \frac{p}{3} \, (\nabla \cdot {\mathbf V_c} )\psi\right]
- \frac{1}{\tau_f}\psi - \frac{1}{\tau_r}\psi\, ,
\label{eq1}
\end{eqnarray}

where $\psi ({\mathbf r},p,t)$ is total phase space density. This equation is valid  for all the types of particles. Isotropic diffusion is defined by the coefficient that depends from rigidity (momentum per
unit of charge, $\rho = \frac{p}{Z}$) $D_{xx} = \beta D_0(\rho/\rho_0)^{\delta}$, inspired by Kolmogorov spectrum ($\delta = 1/3$) of weak magnetohydrodynamic turbulence \cite{SH}. In some models
we have used a break in the index $\delta$ at some rigidity $\rho_0$, with a value $\delta_1 = 0$ below the reference rigidity $\rho_0$. The convection velocity field ${\mathbf V_c}$, that corresponds to the Galactic wind, has a cylindrical symmetry and its z-component is the only one different from zero. It increases linearly with the distance $z$ from the galactic plane, in agreement with magnetohydrodynamical models \cite{Z}. Reacceleration is determined by the diffusion coefficient for the impulse space $D_{pp} $ that is a function of the corresponding configuration space diffusion coefficient and of the Alfven velocity $V_{A}$  in the framework of quasi-linear MHD theory \cite{FSPB}. Of course, Alfven velocity and convection velocity gradient in the Milky Way for reacceleration and convection terms are unknown parameters of propagation (there are no other sources of information from which we could extract them, except the spectra of cosmic rays) and their possible range will be constrained by the analysis of fits of suitable data. The same procedure is valid for constraining the height of the Galactic halo and the other unknown parameters. This will be analyzed further in order to obtain all the possible spectra of antiprotons and positrons using the sets of the constrained parameters. This procedure was already used in \cite{FN} for another propagation code, even for the antiprotons from neutralino annihilations. In the same reference can be found also the review of the propagation models in the Galaxy with an extended colection of historical and recent references. Injected spectra of all primary nuclei are power law in impulse $dq(p)/dp \propto p^{-\gamma}$. This power law approximation has been shown to be allowed in the framework of diffusive shock acceleration models, as well as a small break in the injection index $\gamma$ \cite{BO, SMx, E}. Source term $q({\mathbf r}, p)$ for secondaries contains cross sections for their production from progenitors on H and He targets. The last two terms in equation \ref{eq1} are loss terms with characteristic times for fragmentation and radioactive decay. Propagation equation is solved numerically, using the Crank-Nicholson algorithm.

The heliospheric modulation in the vicinity of the Earth has to be taken in account. We have used a model in which transport equation (that describes diffusion processes in the heliosphere and includes effects of heliospheric magnetic field and Solar wind) is solved in the force field approximation~\cite{GAP}. In this case Solar modulation is a function of just a single parameter that describes the strength of the modulation. All the dynamical processes are simulated with relatively simple changing of the interstellar spectra during the propagation inside the heliosphere, described by the formula

\begin{eqnarray}
\label{eq2}
\frac{ \Phi^{toa} (E^{toa})}{\Phi^{is}(E^{is})} =  (\frac{p^{toa}}{p^{is}})^{2} , \\
E^{is} -  E^{toa} = |Ze| \phi
\end{eqnarray}

where $E$ and $p$ are energies and impulses of the interstellar and top of the atmosphere fluxes and $\phi$ is the unique parameter that determines the solar modulation.

\section{Uncertainties of CR Spectra}

\begin{figure}[ht]
\begin{center}
\includegraphics[width=12.4cm]{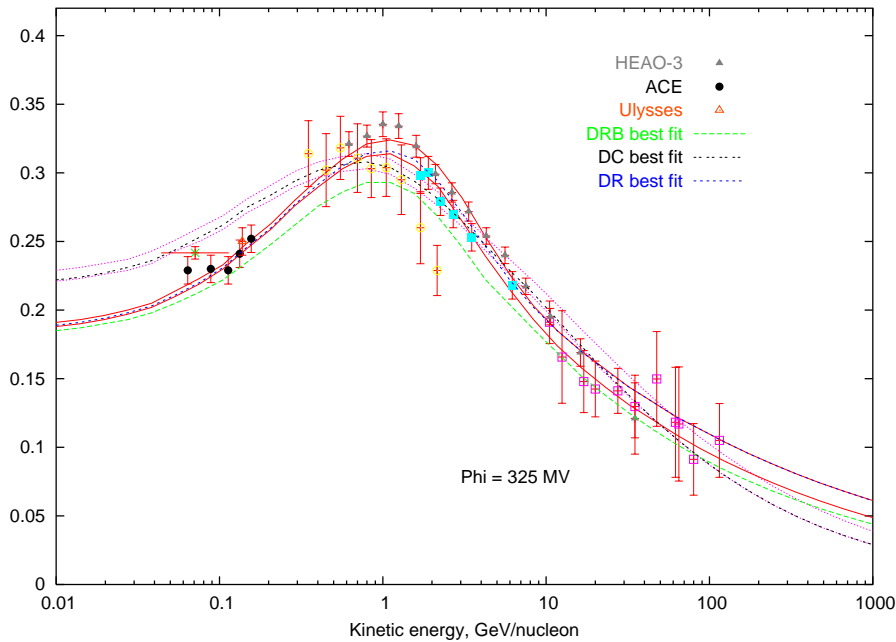}
\end{center}
\caption{Enveloping curves of all the good fits of B/C data for DR and DC model with their best fits inside and the best fit for DRB model. For the complete list of data see \cite{expBC}.}
\label{figBC}
\end{figure}

\begin{figure}[ht]
\begin{center}
\includegraphics[width=7.5cm]{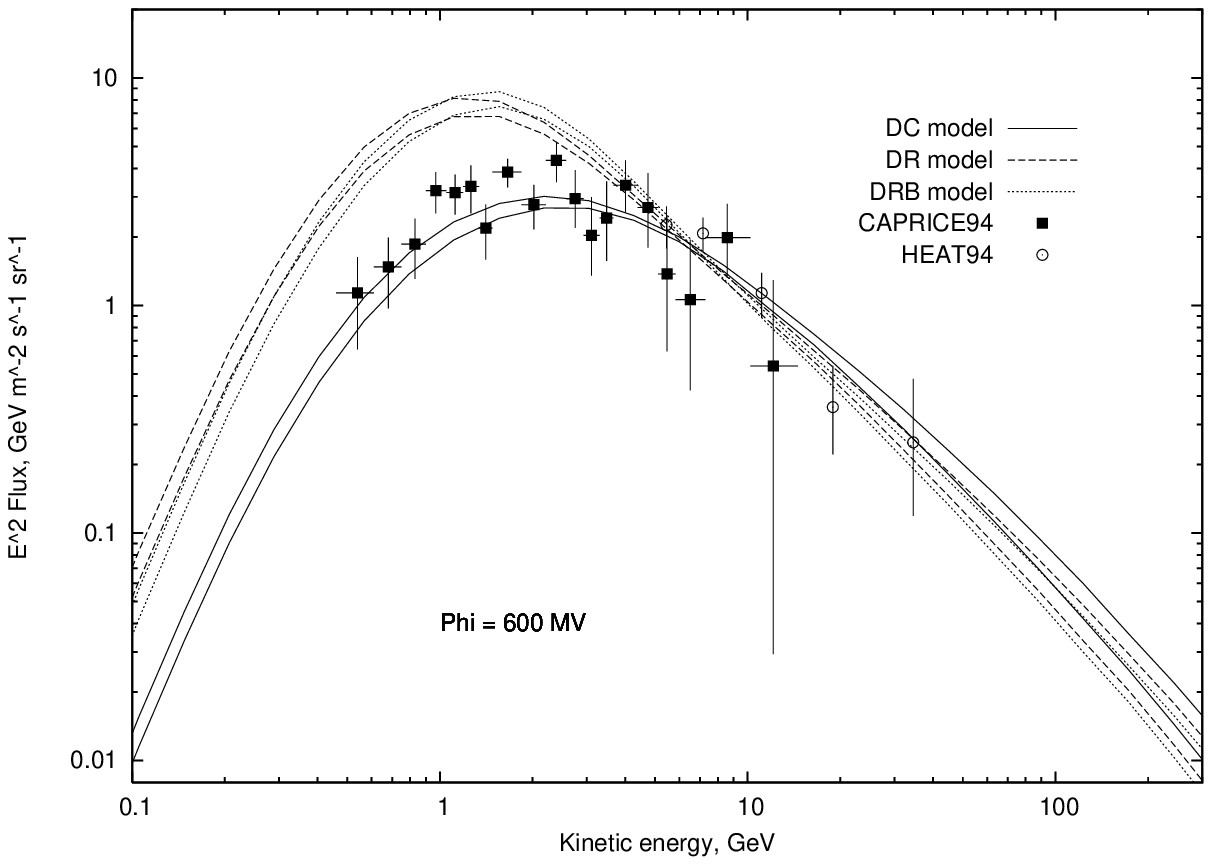}
\includegraphics[width=7.5cm]{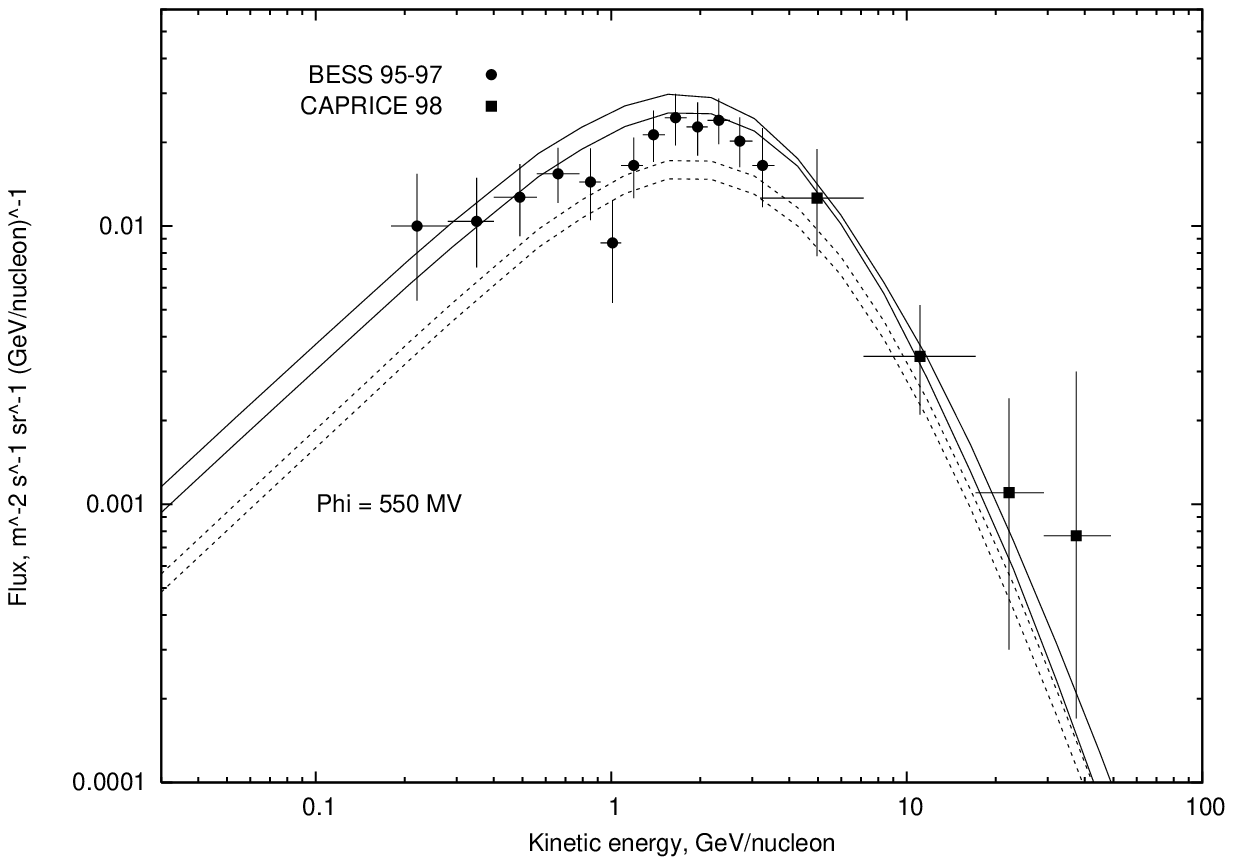}
\end{center}
\caption{Upper and lower bounds of positron (left) and antiproton (right) spectra due to uncertainties of the propagation parameters for DC, DR DRB model and DC and DRB model respectively. Experimental data are taken from \cite{expe, exppbar}.}
\label{fige+pbarparerr}
\end{figure}

We were treating the two extreme cases of propagation models: the
first that uses diffusion and reacceleration (DR) and the second that
contains diffusion and convection (DC) \cite{SM2}.  
Many parameters in the propagation equation are free and must be
constrained by experimental data. 
Secondary to primary CR ratios are the most sensitive quantities on
variation of the propagation parameters. The most accurately measured
parameter is boron to carbon ratio (B/C). We have used a standard $\chi^2$ test:
\begin{equation}
\chi^2 = \sum_{n}
\frac{1}{(\sigma^{B/C_{exp}}_n)^2} (\Phi^{B/C_{exp}}_n -
\Phi^{B/C_{teo}}_n )^2
\end{equation} 
For DR model we have required reduced
$\chi^2$ less than 2 for the fit of the experimental
data~\cite{expBC}~(fig.~\ref{figBC}). 
We take the data with relatively small solar modulation between 325~MV
and 600~MV where the force field approximation is better  justified than for high
modulation parameters. 
The allowed ranges of all parameters are given in table
\ref{tab1}. Using them we have found the enveloping curves of all the $e^+$
and $\bar p$ spectra that present upper and lower bounds of
uncertainty band. For $e^+$ relative uncertainty is varying from 30\%
under 1~GeV to 15\% around 10~GeV (fig.~\ref{fige+pbarparerr}, left) while for
$\bar p$ is from about 10\% up to 15\%. 
For better visibility all the black and white figures that contain
positron and antiproton data will contain just the part of the
experimental data chosen to cover maximally all the energy range,
while all the data are presented on the color figures. 
All data are taken from references \cite{exppbar, expe}.

\begin{table}
\caption{Allowed values for DR model propagation parameters.}
\begin{center}
\renewcommand{\arraystretch}{1}
\setlength\tabcolsep{9pt}
\begin{tabular}{|c|c|c|c|c|c|c|c|}
\hline
par./val. & $z[kpc]$ & $D_{0}[cm^2 s^{-1}]$ & $\delta$ & $\gamma$ & $v_A[kms^{-1}]$ \\ \hline
minimal & 3.0 & 5.2 $10^{28}$ & 0.25 & 2.35 & 22 \\ \hline
best fit & 4.0 & 5.8 $10^{28}$ & 0.29 & 2.47 & 26 \\ \hline
maximal & 5.0 & 6.7 $10^{28}$ & 0.36 & 2.52 & 35 \\ \hline
\end{tabular}
\end{center}
\label{tab1}
\end{table}

\begin{figure}[t]
\begin{center}
\includegraphics[width=7.5cm]{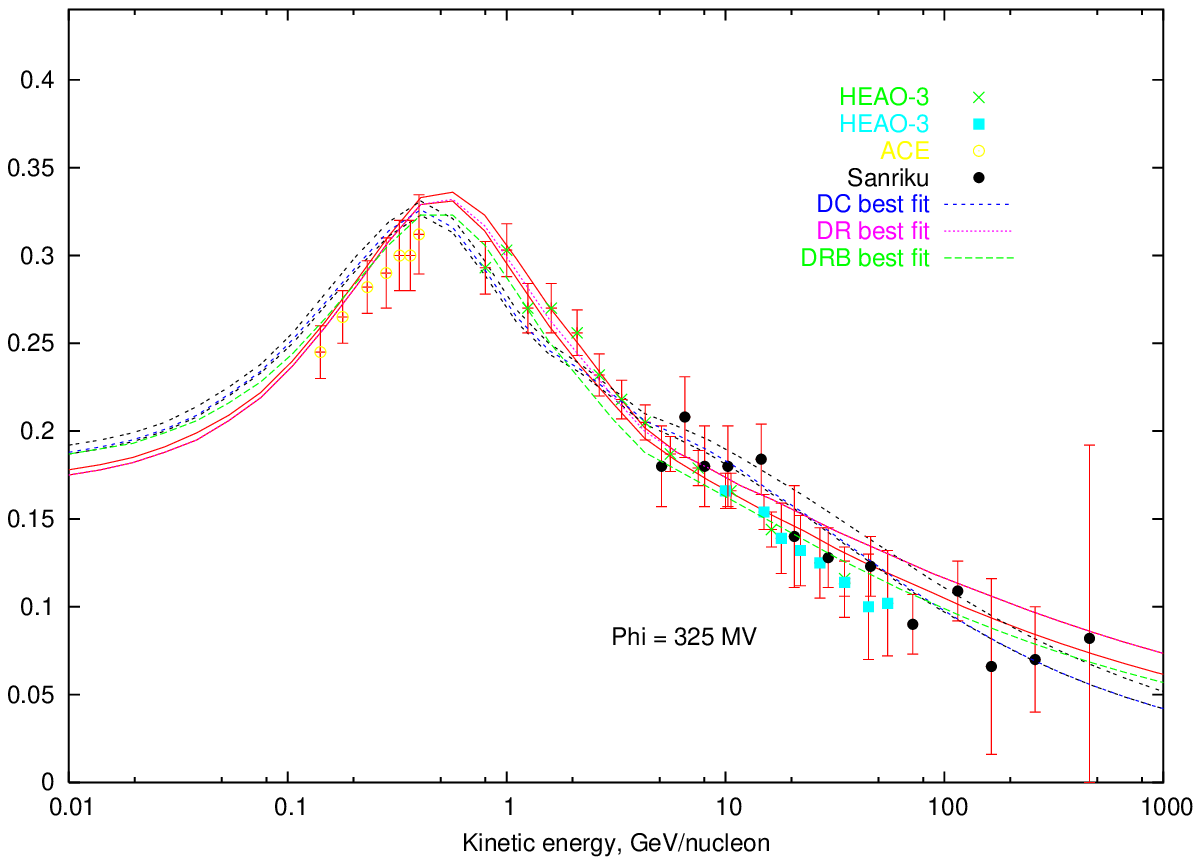}
\includegraphics[width=7.5cm]{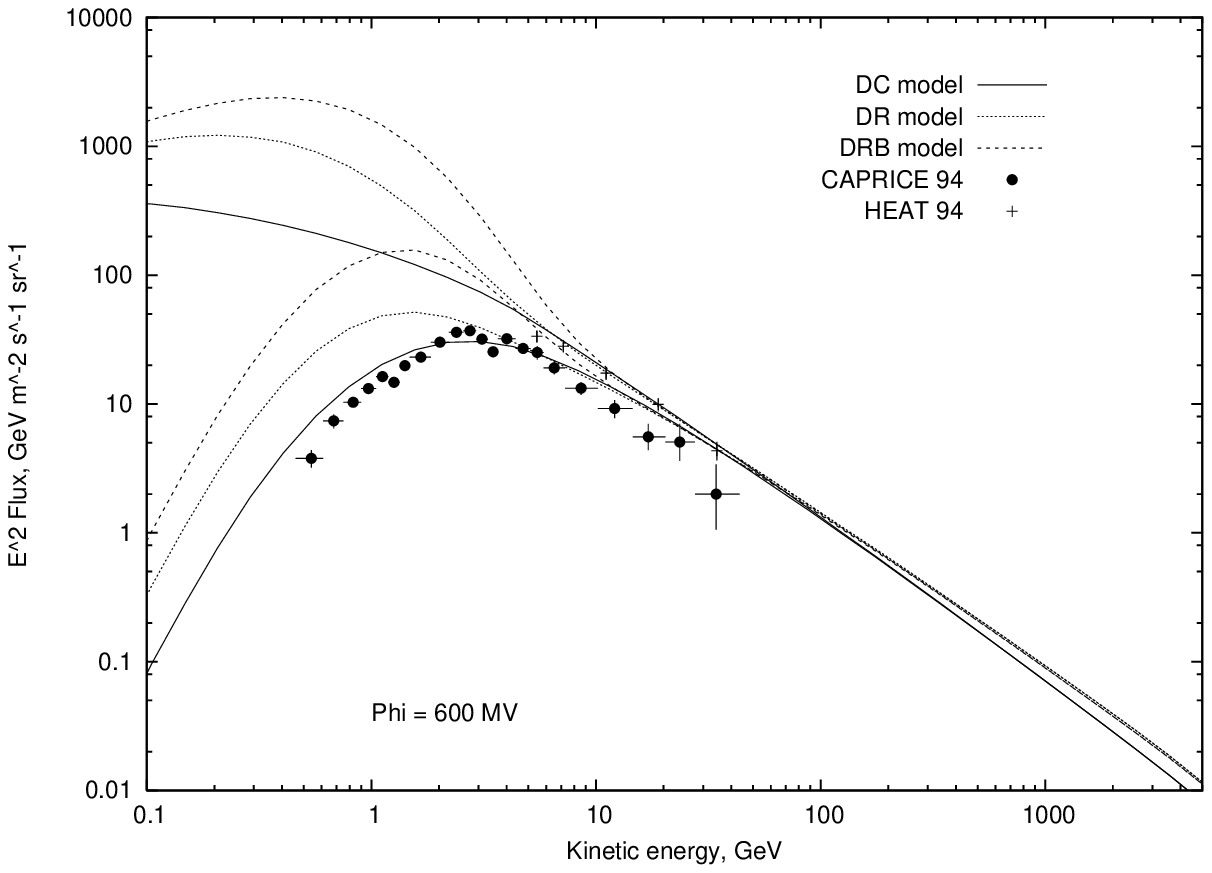}
\end{center}
\caption{Left: Ratios (Sc+Ti+V)/Fe that correspond to the parameters of propagation for DC, DR and DRB model that gave the best fits of B/C data. The upper and the lower limits of all the graphics obtained with the parameters that give good fits of boron to carbon ratio are given for DC and DR model. Experimental data are taken from\cite{expsubFe/Fe}. Right: Top of the atmosphere spectra of $e^-$ that correspond to the parameters of the best B/C fit for DC, DR and DRB model (lower curves) and local interstellar spectra (upper curves). Experimental data are taken from \cite{expe}.}
\label{figFee-}
\end{figure}

\begin{figure}[ht]
\begin{center}
\includegraphics[width=12.3cm]{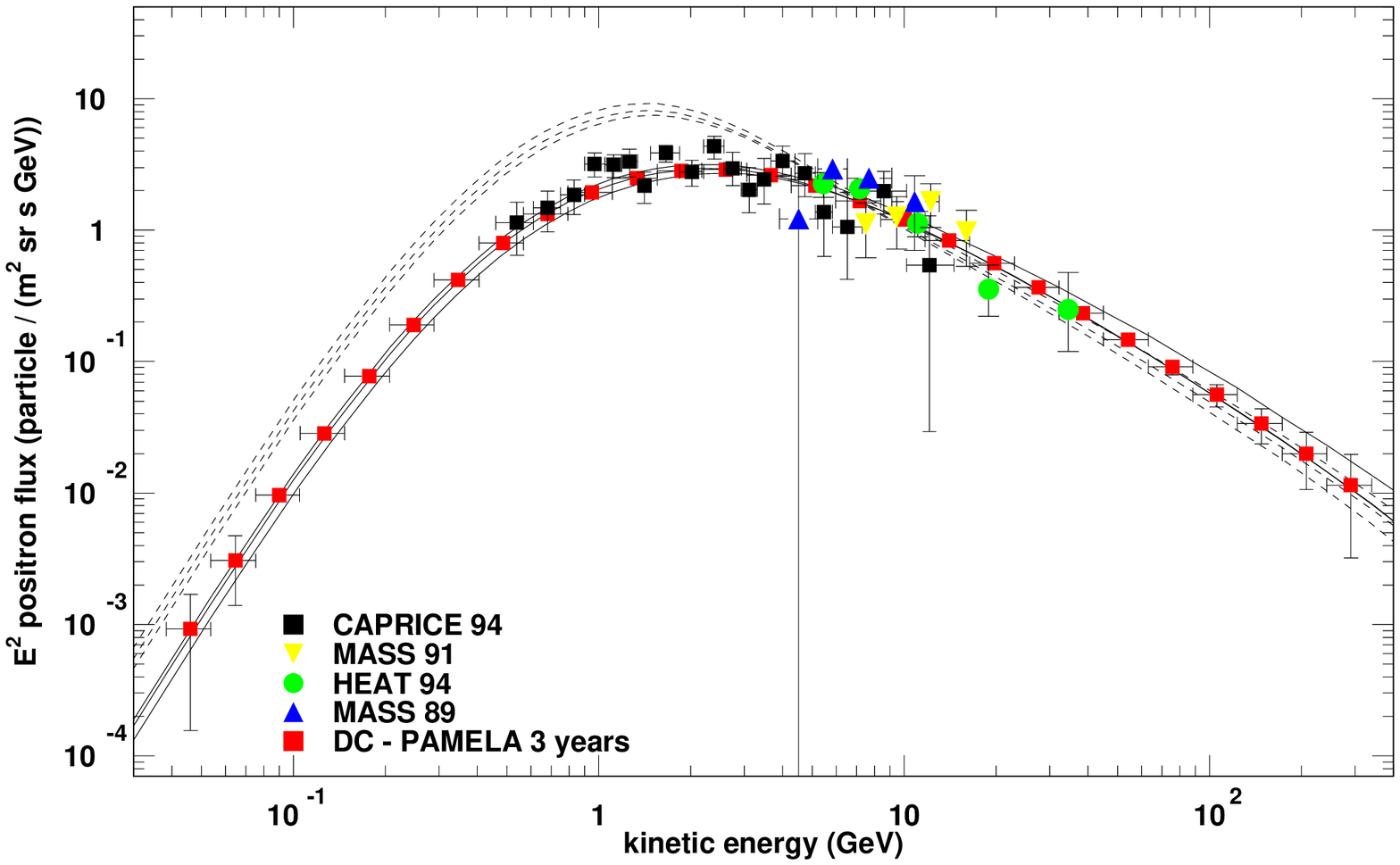}
\end{center}
\caption{Total uncertainties of $e^+$ fluxes and spectra that correspond to the parameters of the best B/C fit for DC (solid lines around the best fit) and DRB model (dashed lines around the best fit). Experimental data (from \cite{expe}) vs.\ PAMELA expectations for DC model.}
\label{fige+}
\vspace{-0.5cm}
\end{figure}

\begin{figure}[ht]
\begin{center}
\includegraphics[width=11.4cm]{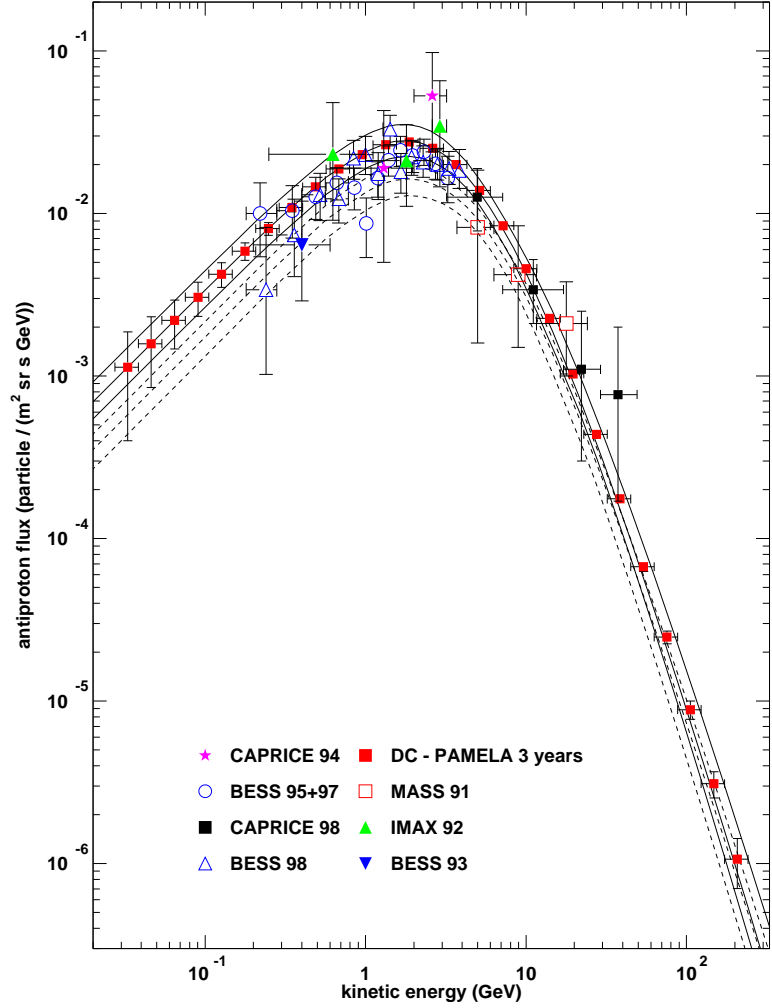}
\end{center}
\caption{Total uncertainty of $\bar p$ fluxes and spectra that correspond to the parameters of the best B/C fit for DC (solid lines) and DRB model (dashed lines). Experimental data (from \cite{exppbar}) vs.\ PAMELA expectations for DC and DRB model.}
\label{figpbar}
\end{figure}

For DC model (fig.~\ref{figBC}) we have taken all the reduced $\chi^2$
values less than 2.8 for the variation of $D_0$, diffusion indexes
$\delta_1$, below, and $\delta_2$, above the reference rigidity
$\rho_0=4$~GV, $z$, $V_c$ and injection index for primary nuclei
$\gamma_1$ below the reference rigidity $\rho^{\gamma}_0=20$~GV and
$\gamma_2$ above it. Positive variations around
$\delta_1=0$ gave unsatisfactory fit. In order to take the smallest
possible break of this index we have decided not to take negative
$\delta_1$ values. Allowed values for the propagation parameters can
be found in table \ref{tab2}. The same analysis as for DR model gives
relative $e^+$ error between 20\% above the maximum and 30\% below it
(fig.~\ref{fige+pbarparerr}, left) while for $\bar p$ is about 20\% around
20~MeV, 17\% around the maximum and 25\% around 20~GeV
(fig.~\ref{fige+pbarparerr}, right). We calculated PAMELA expectations for $e^+$
(fig.~\ref{fige+}) and $\bar p$ (fig.~\ref{figpbar}) (parameters of the
best B/C fit) using its geometrical factor and detector
characteristics \cite{AP} during the three years mission in which it
will measure with high statistics various cosmic rays spectra. We have
found also spectra that correspond to the parameters of the best fit of B/C data for subFe/Fe ratio (another important ratio for testing the parameters of the propagation models, fig.~\ref{figFee-}), protons, He and $e^-$ as well as corresponding uncertainties. For DC model fits are good, while DR overestimates p, He and $e^-$.

\begin{table}
\caption{Allowed values for the propagation parameters for DC model.}
\begin{center}
\renewcommand{\arraystretch}{1}
\setlength\tabcolsep{9pt}
\begin{tabular}{|c|c|c|c|c|c|c|}
\hline
par./val.& $z[kpc]$ & $D_{0}[\frac{cm^2}{s}]$ & $\delta_2$ & $\frac{dV_C}{dz}[\frac{km}{skpc}]$ & $\gamma_1$ & $\gamma_2$ \\ \hline
minimal & 3.0 & 2.3 $10^{28}$ & 0.48 & 5.0 & 2.42 & 2.14 \\ \hline
best fit & 4.0 & 2.5 $10^{28}$ & 0.55 & 6.0 & 2.48 & 2.20 \\ \hline
maximal & 5.0 & 2.7 $10^{28}$ & 0.62 & 7.0 & 2.50 & 2.22 \\ \hline
\end{tabular}
\end{center}
\label{tab2}
\end{table}

We have considered also the DR model with a break in
the injection index for the primary nuclei spectra taken at rigidity
of 10~GV \cite{SM2, E}. We determined allowed values of the
propagation parameters (table \ref{tab3}) demanding the same $\chi^2$
as for DR model (fig.~\ref{figBC}). The positron and antiproton uncertainties are
presented in fig.~\ref{fige+pbarparerr}. Even if $e^+$ are fitted a
little bit better at low energies and also primaries
are fitted better, all of them remain overestimated while $\bar p$
spectra remain practically unchanged. For this reason on
fig.~\ref{figpbar} and fig.~\ref{fige+pbarparerr}, on the right, we refer just the
DRB case.

\begin{table}
\caption{Allowed values for the propagation parameters for DR model with the break.}
\begin{center}
\renewcommand{\arraystretch}{1}
\setlength\tabcolsep{9pt}
\begin{tabular}{|c|c|c|c|c|c|c|c|}
\hline
par./val. & $z[kpc]$ & $D_{0}[\frac{cm^2}{s}]$ & $\delta$ & $\gamma_1$ & $\gamma_2$ & $v_{A}[\frac{km}{s}]$ \\ \hline
minimal & 3.5 & 5.9 $10^{28}$ & 0.28 & 1.88 & 2.36 & 25 \\ \hline
best fit & 4.0 & 6.1 $10^{28}$ & 0.34 & 1.92 & 2.42 & 32 \\ \hline
maximal & 4.5 & 6.3 $10^{28}$ & 0.36 & 2.02 & 2.50 & 33 \\ \hline
\end{tabular}
\end{center}
\label{tab3}
\end{table}

We have also seen how the obtained antiproton spectra
change on variation of the most important antiproton production cross
sections. Those are reactions that include all the types of hydrogen
and helium. Antiprotons are created in the interactions of primary
cosmic rays (protons and other nuclei) of sufficiently high energies
with interstellar gas. Dominant processes are interactions of high
energy primary protons with hydrogen, for example $p + p \rightarrow p
+ p + p + \bar p $. Parameterization of this cross sections used in our
version of Galprop code is given by Tan and Ng \cite{TN}. Other used
cross sections, those of primary protons with other nuclei, are
studied in reference \cite{GS}. From these, the most important are those
that involve helium and they contribute less than 20\% of the total
production of all the antiprotons. All the heavier nuclei together
give just a few percents of the total production. Simultaneous
settings of all the production cross sections to the maximum/minimum
rise/lower the upper/lower parameter uncertainty bounds. Errors
obtained in this way give contributions to the total uncertainties
varying from 20\% up to 25\% in the case of DR model and, almost the
same, from 20\% up to 24\% for DC model, depending of the energy range
of the spectra. Non production cross sections, the so called tertiary component, correspond to inelastically scattered secondaries $\bar p + X \rightarrow \bar p + \tilde X$. Those processes tend to bring down the energies of the antiprotons of relatively high energies, flattening like that the spectra. But, even if their uncertainty is relatively big, this does not give relevant change of the spectra because tertiary contribution is very small.

Changing of another exactly unknown quantity, He/H ratio, in a reasonable range from 0.08 to 0.11 gives a relatively small contribution, that vary from 3\% - 7\% depending of the energy, for both $e^+$ and $\bar p$ uncertainty, and for both of the models. Total uncertainties of $e^+$ and $\bar p$ are presented at fig.~\ref{fige+} and fig.~\ref{figpbar} respectively. They vary from 35\% up to 55\% for antiprotons and from 20\% up to 40\% for positrons roughly for both of the models in the current experimental data energy range.

\section{Antiproton Flux from Neutralino Annihilations}
In this section we take into account the possibility of a
neutralino induced component in the $\bar{p}$ flux. 
Our analysis is performed in the well known mSUGRA
framework~\cite{msugra} with the usual gaugino mass universality at the
grand unification scale $M_{GUT}$. The five input parameters are
defined as follows:
\[ m_{1/2},\;\;  m_0,\;\;  sign(\mu),\;\;  A_0\;\; \rm{and}\;\; \tan\beta \;,
\]
where $m_0$ is the common scalar mass, $m_{1/2}$ is the common gaugino mass and $A_0$ 
is the 
proportionality factor between the supersymmetry breaking trilinear couplings and the 
Yukawa couplings. $\tan\beta$ denotes the ratio of the vacuum
expectation values of the two neutral components of the SU(2) Higgs doublet, while
the Higgs mixing $\mu$ is determined (up to a sign) by imposing the Electro-Weak Symmetry 
Breaking (EWSB) conditions at the weak scale. The parameters at the weak energy scale 
are determined by the evolution of those at the unification scale, according to the renormalization 
group equations (RGEs)~\cite{Cesarini:2003nr}. For this purpose, we have made use of the ISASUGRA RGE package 
in the ISAJET 7.64 software~\cite{isasugra}. After fixing the five mSUGRA parameters at the unification scale, we extract from the ISASUGRA output
the weak-scale supersymmetric mass spectrum and the relative mixings. Cases in which the
lightest neutralino is not the lightest supersymmetric particle or there is no
radiative EWSB  are disregarded.
We have assumed a small clump scenario~\cite{Bergstrom:1998jj} for the dark matter halo in
our galaxy. By hypothesis the clump is a spherical symmetric compact object with
mass $M_{cl}$ and density profile
$\rho_{cl}\left(\vec{r}_{cl}\right)$. 
We denote with $f$ the dark matter fraction concentrated
in clumps and we introduce the dimensionless parameter $d$
\begin{equation}
d=\frac{1}{\rho_0}\frac{\int d^3 r_{\rm cl}\left[\rho_{\rm
      cl}\left(\vec{r}_{\rm cl}\right)\right]^2}{\int d^3 {
      r_{\rm cl}}\rho_{\rm cl}\left(\vec{r}_{\rm cl}\right)}
\label{eq:pbarsusyflux}
\end{equation}  
that gives the overdensity due to a clump with respect to the local
halo density
%
%
%
%
%
 $\rho_0=\rho(r_0)$, where $r_0$ is our distance from the
Galactic center. 
In a smooth halo scenario the total neutralino induced $\bar{p}$ flux
calculated for $r=r_o$ is given by~\cite{Bergstrom:1999jc} 
\begin{equation}
  \Phi_{\bar{p}}(r_0,T) \equiv
  (\sigma_{\rm ann}v) \sum_{f}^{}\frac{dN^{f}}{dT}B^{f}
  \left(\frac{\rho_0}{m_{\tilde{\chi}}}\right)^{2}
  \, C_{\rm prop}(T) \;.
\end{equation}
where $T$ is the $\bar{p}$ kinetic energy, $\sigma_{\rm ann}v$ is the
total annihilation cross section times the relative velocity, $m_\chi$ is the
neutralino mass, $B^{f}$ and $dN^{f}/dT$, respectively, the
branching ratio and the number of $\bar{p}$ produced in each
annihilation channel $f$ per unit energy and $C_{\rm prop}(T)$ is a function entirely determined by the
propagation model. 

In the presence of many small clumps the $\bar{p}$
flux is given by
\begin{equation}
\Phi^{\rm clumpy}_{\bar{p}}(r_0,T)=fd\cdot \Phi_{\bar{p}}(r_0,T)
\end{equation}
For the smooth profile we have assumed a Navarro, Frenck and White profile (NFW)~\cite{navarro}.

The primary contribution to the $\bar{p}$ flux has been computed
using the public code DarkSUSY~\cite{Gondolo:2004sc}. We have modified
the $\bar{p}$ propagation in order to be
consistent with the DC model as implemented in Galprop. We assumed diffusion coefficient spectra
used in Galprop with our best fit values for the
diffusion constants $D_0$ and $\delta$.
In DarkSUSY the convection velocity field is constant in the upper and
lower Galactic hemispheres (with opposite signs, and so it suffers
unnatural discontinuity in the Galactic plane) while Galprop uses
magnetohydrodynamically induced model in which component of velocity field along the Galactic latitude (the only one different from zero) increases
linearly with the galactic latitude~\cite{Z}. We have assumed an
averaged convection velocity calculated from the Galactic plane up to
the Galactic halo height $z$.

We present here four different susy contribution to the $\bar{p}$
flux for different neutralino masses (obtained from a particular
choice of mSUGRA parameters) and different clumpiness factor $fd$. 
 Higher neutralino masses  improve high energy data fit but only with the 
increase of the clumpiness factor because of the dependence from the inverse
neutralino mass squared $m_\chi$ in the $\bar{p}$ flux formula~(\ref{eq:pbarsusyflux}).
We remark that for those models we have not
computed the neutralino relic density.  

\begin{figure}[ht!]
\begin{center}
\includegraphics[width=7.5cm]{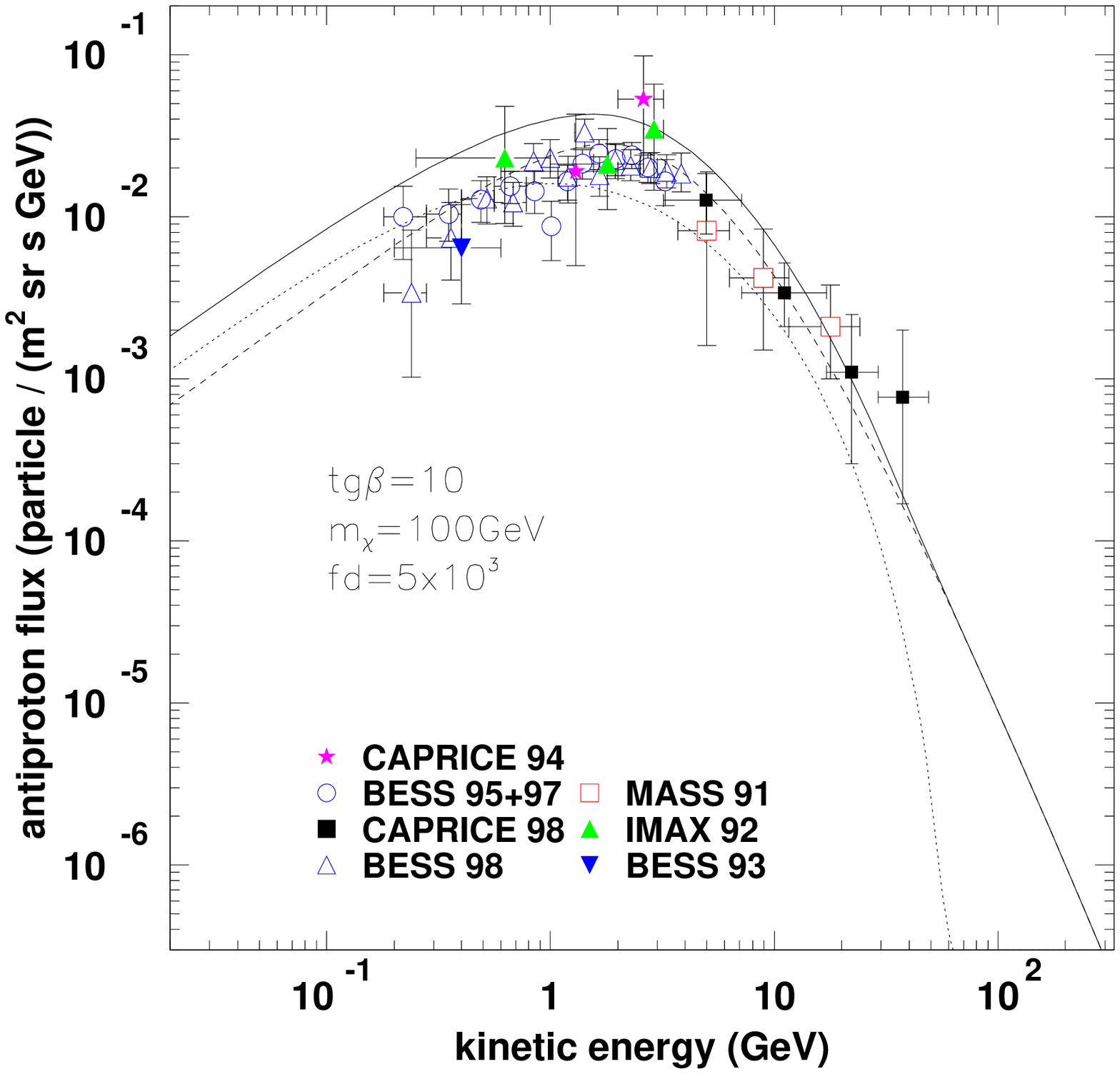}
\includegraphics[width=7.5cm]{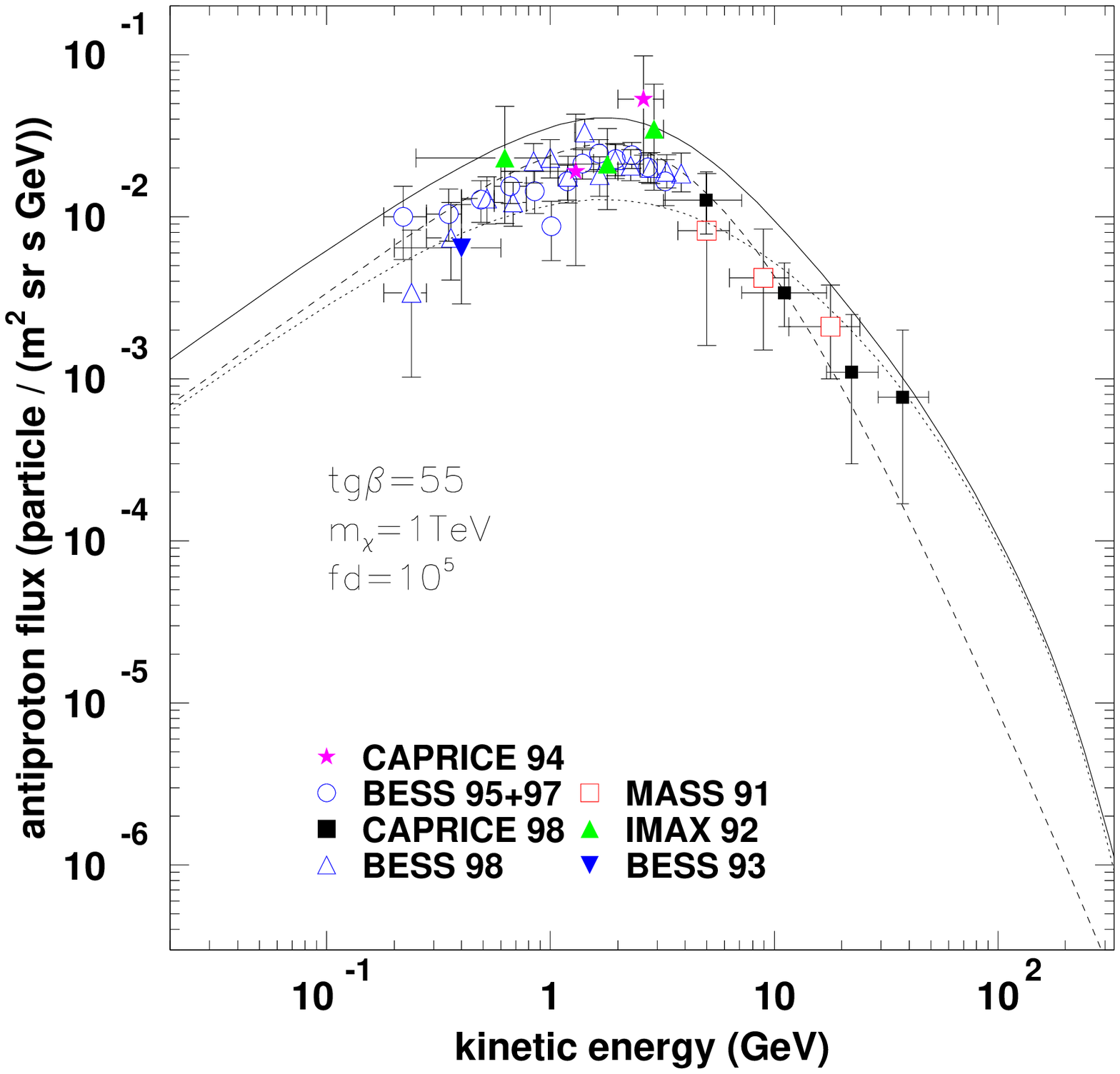}
\includegraphics[width=7.5cm]{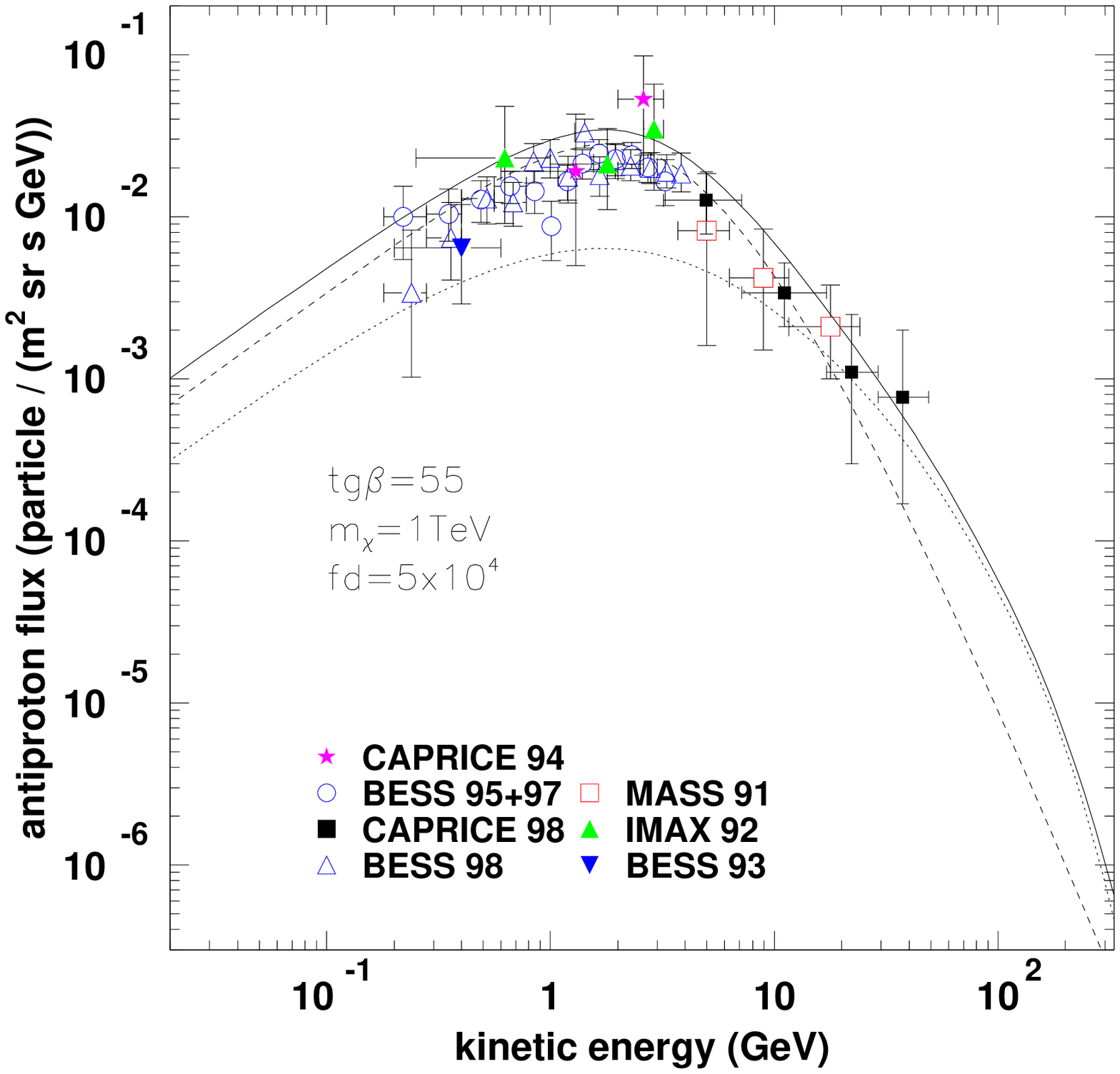}
\includegraphics[width=7.5cm]{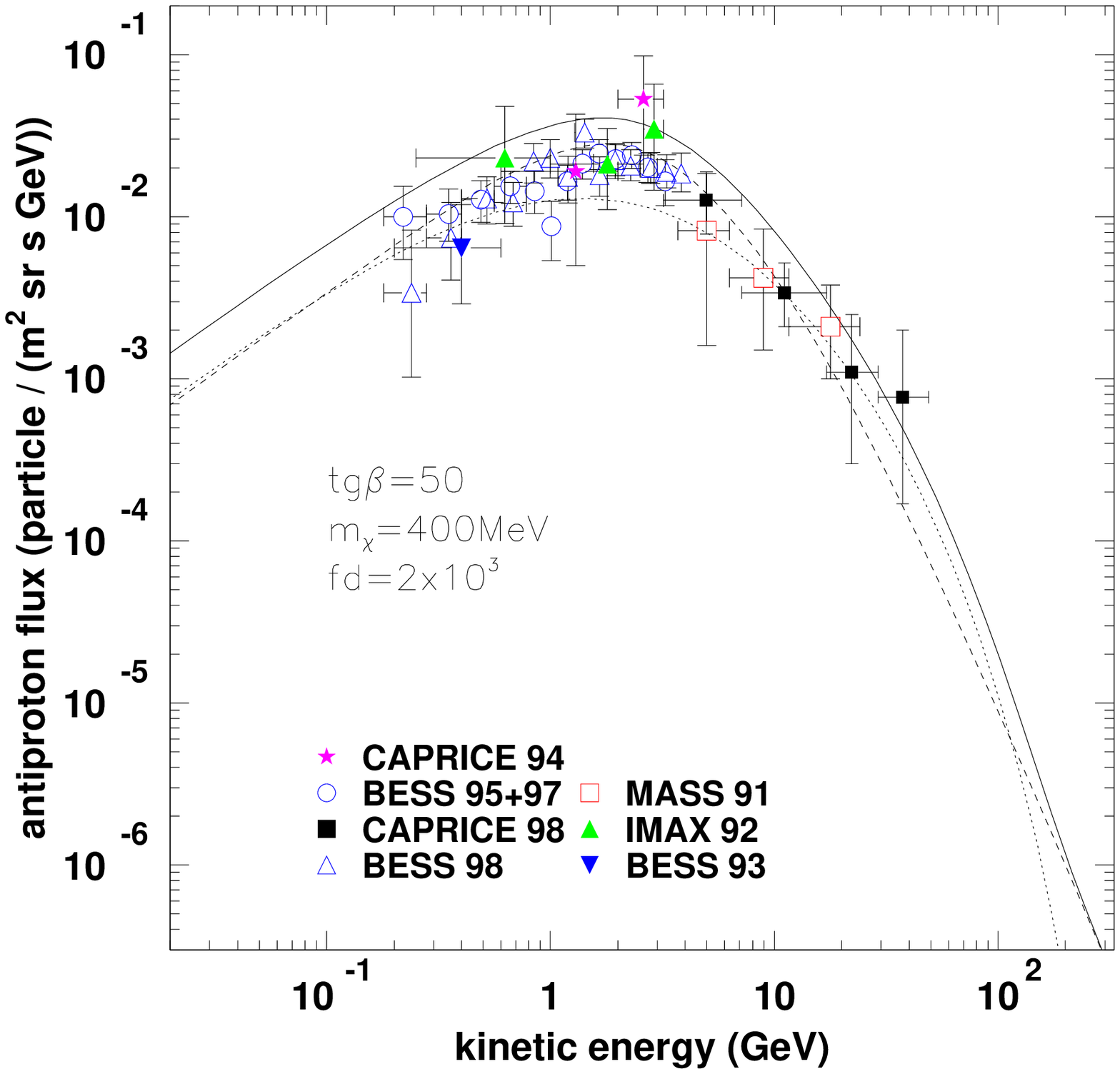}
\caption{Different susy contribution to the total $\bar{p}$ flux with
the DC model background. The dashed lines correspond to
the background contribution, the punctuated lines correspond to the
neutralino induced contribution while the solid lines correspond to
total contribution}
\end{center}
\end{figure}

\section{Conclusions}

For $e^+$ in DR model, even when the uncertainties are included, the
curve of the minimal $e^+$ production still remains above the
experimental results. Breaking the primary spectra gently improves
just the low energy part of the spectra below the maximum, but even
the minimal predictions still remain bigger than the experimental
results around the maximum as well as below it. On the other side,
this slightly changes the best B/C fit (fig.~\ref{figBC}). This is
also not sufficient to match protons and helium, that are still
overestimated. In the end, $e^-$ remain largely overproduced at low
energies, even more than without the break. Uncertainty bands of $\bar
p$ in DR models touch the experimental data from below. In this case
the fit can be
improved easily with any primary component that is coming from
eventual neutralino annihilation or some other exotic
contribution. For DC model all the results are in good agreement with
the data, with only some problems with B/C data.

As a possibility for the future considerations we would like to
emphasize that it is natural to expect models that include both of the
processes -- convection and reacceleration, and this perhaps can be a
key to solve simultaneously problems with positrons, electrons and
B/C. But, we have not yet found such a kind of models. 
In any case, further measurements of $\bar p$ and $e^+$ spectra, primary to secondary CR ratios and Solar effects on them, as well as the precise determination of important nuclear cross sections seem to be crucial for answering the question of the models quality.

In the framework of DC model exotic contributions remain possible at high
energies (E $>$ 20 GeV), and also they are not excluded at lower energies
due to the relatively large uncertainties (if the Standard Model
productions are smaller than the best fits). 
In the case of DR model background it is possible to improve the
$\bar{p}$ data fit adding a neutralino induced component for a large
part of the parameter space.
Finally we have shown how a possible neutralino induced component in a clumpy halo
scenario could give a contribution to the total $\bar{p}$ flux, in the
case of DC model background.

\end{document}